\documentclass[a4paper,11pt]{amsart}
\usepackage{multicol,ifthen}
\usepackage{amssymb,stmaryrd}
\usepackage{harvard}
\usepackage{graphicx}

\newcommand{\beq}{\begin{equation}}
\newcommand{\eeq}{\end{equation}}
\newcommand{\beqarr}{\begin{eqnarray}}
\newcommand{\eeqarr}{\end{eqnarray}}
\newcommand{\beqa}{\begin{eqnarray*}}
\newcommand{\eeqa}{\end{eqnarray*}}
\begin{document}

\title{Correction note to\\  {\em Another look at the Pioneer anomaly }  }
\author[E. Scholz]{Erhard Scholz$^{\, \natural}$} \thanks{ $\natural$  
scholz@math.uni-wuppertal.de, University Wuppertal,   \\
\hspace*{7mm} Department C, Mathematics and Natural Sciences, and \\
\hspace*{6.5mm} Interdisciplinary Center for Science and Technology Studies}

\date{August 14, 2007 }

\begin{abstract}
The often quoted equ. (15) in \cite{Anderson_ea:2002} defines the observed   Pioneer anomaly $a_P$  such that  literal reading suggests  $ \nu _{obs} < \nu_{mod}  $ ($ \nu _{obs} $ the observed (Doppler) frequency, $\nu_{mod}$  the one expected from a relativistic orbit model of the spacecraft). Contextualized reading shows, however,  that (15)   is meant the other way round, $ \nu _{obs} > \nu_{mod}  $. Version 1 to 4 of  
\cite{Scholz:Pioneer}  built upon the first reading and led to a correction term for frequency shifts of wrong sign. Thus the proposal therein  does not resolve the Pioneer anomaly, although it gives a correction term for frequencies which has the correct absolute value.
\end{abstract}

\maketitle 

\section{On signs}
The often quoted equ. (15) in \cite{Anderson_ea:2002} (1st line) defines the observed   Pioneer anomaly $a_P$ by
\setcounter{equation}{14}
\beq [\nu _{obs}(t) - \nu_{mod}(t)]_{DSN}  = - \nu _0\,\frac{2 a_p t}{c} \; ,\eeq
where DSN (deep space network) stands for JPL's (Jet Propulsion Laboratory's)  network of ground stations for tracking and navigating interplanetary satellites.  $\nu _{obs} $ denotes ``the frequency of the transmitted signal observed by a DSN antenna, and $\nu_{model}$[,] the predicted frequency of the signal'', to quote  the authors  (ibid., p. 17). $a_P$ is an anomalous (unmodelled) acceleration term of positive sign, $t$ the time difference to the beginning of an interval of model calculation. Referring to their endnote [38], the authors warn  that the DSN convention for the Doppler frequency shift is
\[ (\Delta \nu )_{DSN}:= \nu _0-\nu \;, \]
(``where $\nu $ is the measured frequency and $\nu _0$ ist the reference frequency'')  while the ``usual'' convention for Doppler shift  works with
\[  (\Delta \nu )_{usual}:= \nu -\nu_0 \; .  \]

As the DSN convention only applies to $\triangle \nu $'s (Doppler {\em shifts}), not to frequencies themselves, equation (15)  remains 
\[ \hspace{-75mm}  \hspace{25mm}   (\dag) \hspace{20mm} \quad \quad   \nu _{obs} - \nu_{mod} = - \nu _0\,\frac{2 a_p t}{c} \; ,  \]
which of course implies $ \nu _{obs} < \nu_{mod}  $. 
For readers who wonder whether the  warning may be meant to refer  to the indexed square bracketed term in equ. (15) also, the authors give an additional explanation at the end of the  paragraph: 
\begin{quote}
``By DSN convention [38], the first of Eqs. (15) is
\[  [\Delta \nu _{obs} - \Delta \nu_{mod}]_{usual} = -  [\Delta \nu _{obs} - \Delta \nu_{mod}]_{DSN}  \mbox{.'' } \;\; \]
\cite[17]{Anderson_ea:2002}
\end{quote}
Because of
 \[ -  [\Delta \nu _{obs} - \Delta \nu_{mod}]_{DSN} =  -\left( (\nu _0- \nu _{obs}) - (\nu _0-\nu_{mod})\right) =\nu _{obs} - \nu_{mod} \; ,\]
this is again  (\dag ).
Accordingly, this reading has been used in  \cite[v1--v4]{Scholz:Pioneer}.
  But there are contextual problems.

The authors of \cite{Anderson_ea:2002} characterize the Pioneer anomaly as ``a slight blueshift on top of the larger redshift'' (ibid., 17).   \cite[v1--v4]{Scholz:Pioneer} has taken equ. (15) and the succeeding eplanation as an authoritative mathematical description of the Pioneer anomaly and proposed a (twisted) interpretation of its characterization as a blueshift by the necessity to subtract a correction term. But that seems to be {\em wrong}.

Other parts of the paper  \cite{Anderson_ea:2002} show  that the authors want (15) to be understood  in the sense of  
\[ \hspace{-70mm}  \hspace{25mm}   (\ast) \hspace{20mm} \quad \quad   
\nu_{mod} - \nu _{obs}   = - \nu _0\,\frac{2 a_p t}{c} < 0 \;    \]
and  $ \nu_{mod} < \nu _{obs}$.  This is most clearly expressed in the specification that  the ``Doppler residuals (observed Doppler velocity minus model Doppler velocity)'' is {\em negative} (figure 8). In terms of DSN convention, this  is ($\ast$) (observed redshift  smaller than the redshift expected from the model calculations).

There seems to be a wide  consense in the Pioneer literature that equ. 
(15) has to be understood in the sense of   $(\ast)$. I thank H.-J. Fahr,  R. Plaga and F. Steinle to have helped me  accepting that this is the contextually correct(ed) reading. 

I only have to add that then  (15)  seems to have  omitted $\Delta $'s and should read as
\[ \hspace{-60mm}  \hspace{25mm}   (15') \hspace{20mm} \quad \quad 
[ \Delta  \nu _{obs} - \Delta \nu_{mod} ]_{DSN} = - \nu _0\,\frac{2 a_p t}{c} \; . \]
Then the l.h.s. corresponds to the verbal description in the caption of figure 8,  
although now the explanation in the paragraph following the equation in \cite{Anderson_ea:2002}  carries a {\em sign error}. It  has to be changed to: ``By DSN convention [38], the (l.h.s. of the) first of Eqs. (15) is
$- [\Delta \nu _{obs} - \Delta \nu_{mod}]_{usual} =   [\Delta \nu _{obs} - \Delta \nu_{mod}]_{DSN}$.''

\section{Withdrawal of the proposal made in (Scholz 2007) }
In the light of the corrected sign in equ. (15) \cite{Anderson_ea:2002}, the attempted explanation of the Pioneer anomaly in \cite[v1--v4]{Scholz:Pioneer}  leads to a correction term of wrong sign.\footnote{Apologies to J. Masreliez with respect to the sign criticism in \cite[footnote (2)]{Scholz:Pioneer}. }
 It {\em cannot be upheld} and has to be {\em  withdrawn}. 

It remains true that under the assumption of a  Weyl geometric explanation of the Hubble effect,  a frequency correction has to be applied to the raw data, in order to isolate the pure Doppler signal from the whole redshift.  In contrast to the standard model of cosmology this leads to a correction at the order of magnitude of $a_P$.
But assuming ($\ast$), the observed anomaly would appear to be twice the one given  in the present literature. Therefore  $a_P$ cannot be explained in this way as an immediate result of the Hubble effect. Secondary effects of the Hubble correction term for velocities would arise from the relativistic (PPN) orbit equation  \cite[equ. (3)]{Anderson_ea:2002}, but seem to be of smaller order of magnitude.

\end{document}